\documentclass[twocolumn]{aastex62}
\newcommand{\HI}{\hbox{{\rm H}\kern 0.2em{\sc i}}}
\newcommand{\Msun}{$M_{\odot}$}
\shorttitle{Extraplanar Molecular Gas in a Virgo Spiral}
\shortauthors{Lee \& Chung}

\begin{document}

\title{The ALMA Detection of Extraplanar $^{13}$CO in a Ram-pressure-stripped Galaxy and Its Implication}

\correspondingauthor{Aeree Chung}
\email{bhlee301@gmail.com, achung@yonsei.ac.kr}

\author[0000-0002-3810-1806]{Bumhyun Lee}
\affiliation{Department of Astronomy, Yonsei University, 50 Yonsei-ro, Seodaemun-gu, Seoul 03722, Republic of Korea}

\author[0000-0003-1440-8552]{Aeree Chung}
\affiliation{Department of Astronomy, Yonsei University, 50 Yonsei-ro, Seodaemun-gu, Seoul 03722, Republic of Korea}




\begin{abstract}
NGC~4522 is a Virgo spiral that is currently undergoing active ram pressure stripping. In previous single-dish observations, $^{12}$CO emission was detected outside of the stellar disk, some of which coincides with the extraplanar {\HI} gas and H$\alpha$ patches. The extraplanar gas identified in multi-wavelength data makes this galaxy an ideal case to study the impact of pressure due to the cluster medium on the interstellar gas of various phases. In this Letter, we present the high-resolution $^{12}$CO(1--0) and $^{13}$CO(1--0) data of NGC~4522 obtained using the Atacama Large Millimeter/submillimeter Array (ALMA). In particular, we report here the extraplanar $^{13}$CO detection that has never before been seen in ram-pressure-stripped galaxies. As the main donor of $^{13}$C in the interstellar medium is evolved stars, the presence of $^{13}$CO strongly suggests that heavy elements likely originated from the galactic disk, not from the newly formed stars in situ. Even though it is still inconclusive whether it is stripped in atomic form or as molecules, this study provides evidence for the ram pressure stripping of heavy elements, which can chemically enrich the halo gas, and potentially the intracluster medium, in the case that they are pushed strongly enough to escape the galaxy.
\end{abstract}




\section{INTRODUCTION} \label{sec:intro}
Galaxies in dense environments can be ram pressure stripped by the intracluster medium (ICM), losing interstellar gas \citep{gunn72}. Although it is evident that diffuse interstellar gasses like {\HI} can be quite easily removed from galaxies by dynamical pressure of the ICM \citep[e.g.,][]{chung09}, whether denser (hence, star-forming) molecular gas can be also stripped by the cluster medium is rather unclear and is still a matter of debate \citep{kenney89,boselli14,chung17}. Consequently, the detailed process of star formation quenching after the {\HI} stripping of galaxies in the cluster environment is not yet fully understood.

To obtain a more complete understanding of the effect of ram pressure on molecular gas in galaxies, we recently carried out a detailed study of selected Virgo spirals that are undergoing active {\HI} stripping events. Using high-resolution CO data, we showed that the overall CO morphology shares many characteristics with {\HI}, which indicates that the dense molecular gas is also influenced by ram pressure in similar ways as diffuse atomic gas \citep{lee17}. However, in our study, we did not see any molecular gas outside of the stellar disk, as in some previous single-dish observations \citep[e.g.,][]{vollmer08}. Hence, it remains unclear as to whether or not molecular gas could be directly removed from the disk or formed from stripped atomic gas, as well as what happens to the extraplanar molecular gas in either case. 

From this point of view NGC~4522, located in the Virgo cluster, is particularly interesting as it is the nearest example among only a few ram-pressure-stripped cases where extraplanar molecular gas has been identified  \citep[e.g.,][]{vollmer08, jachym14, jachym17, wong14, verdugo15}. With 30\% of {\HI} and 15\% of $^{12}$CO (2--1) found outside of the stellar disk (Table~\ref{tab:table1}), it provides an ideal laboratory to study the influence of ram pressure on the molecular gas in depth.

In this Letter, we present $^{12}$CO(1--0) and $^{13}$CO(1--0) data of NGC~4522 obtained using the Atacama Large Millimeter/submillimeter Array (ALMA). In particular, we report for the first time the extraplanar $^{13}$CO emission associated with a ram-pressure-stripped galaxy. Using the ALMA data, we probe the origin of the extraplanar molecular gas and discuss its impact on the surrounding medium of the host galaxy.

\begin{deluxetable}{lcc}[h]
\tablecaption{General properties of NGC~4522 \label{tab:table1}}
\tablewidth{0pt}
\tablehead{ \colhead{Parameter} & \colhead{Value} }

\startdata
R.A. (J2000)$^{a}$ & $12^h33^m39^s.71$\\
Decl. (J2000)$^{a}$ & $+09\degr10\arcmin29\arcsec.7$\\
Morphological type$^{a}$		& SBc\\
Inclination ($\degr$)$^{a}$ & 79.2\\
Position angle ($\degr$)$^{a}$ & 34.5\\
$D_{25}$ (arcmin)$^{a}$ & 3.47\\
$def_{\rm\tiny {\HI}}$$^{b}$ & 0.86\\
$v_{\rm\tiny {\HI}}$ (km~s$^{-1}$)$^{b}$ & 2331\\
Main stellar disk &  & \\
~~~$M_{\rm\tiny {\HI}}$~(\Msun)$^{c}$ & 2.3 $\times$ 10$^8$\\
~~~$M_{\rm {H_{2}}}$~(\Msun)$^{c}$ & 1.7 $\times$ 10$^8$\\
Extraplanar region &  & \\
~~~$M_{\rm\tiny {\HI}}$~~(\Msun)$^{c}$ & 1.0 $\times$ 10$^8$\\
~~~$M_{\rm {H_{2}}}$~(\Msun)$^{c}$ & 0.3 $\times$ 10$^8$\\
\enddata
\tablecomments{(a) HyperLeda \citep[][\url{http://leda.univ-lyon1.fr/}]{makarov14}; (b) \cite{chung09}; (c) Measured inside and outside of the optical disk (blue ellipse, Figure~\ref{fig:fig1}) using data from \cite{chung09} and \cite{vollmer08}.}
\end{deluxetable}


\section{ALMA OBSERVATIONS} \label{sec:obs}
Our ALMA observations (project ID: 2015.1.01341.S; PI: B. Lee) were carried out during Cycle 3 (2016 January) using the 12m array. Two regions were observed in both $^{12}$CO(1--0) ($\nu_{\rm rest} = 115.271~$GHz) and $^{13}$CO(1--0) ($\nu_{\rm rest} = 110.201~$GHz) lines (Figure~\ref{fig:fig1}). The first pointing was placed at the galactic center, and the second pointing was centered around the local peak of the extraplanar $^{12}$CO gas \citep{vollmer08}, covering most of the extraplanar {\HI} and H$\alpha$ emission in the southwest \citep{kenney04}. The primary beam at these frequencies is $\sim$52$\arcsec$, which is comparable to $\sim$1/3 of NGC~4522's stellar disk. In total, 51 minutes and 273 minutes were integrated on the target for $^{12}$CO and $^{13}$CO with 46 and 43 antennas, respectively, yielding 1.44~mJy~beam$^{-1}$ and 0.46~mJy~beam$^{-1}$ over a 5~km~s$^{-1}$ channel width. The data were calibrated using the ALMA pipeline in Common Astronomy Software Applications (CASA)-4.5.3 \citep{mcmullin07}. For imaging, the data were naturally weighted to obtain the highest signal-to-noise ratio (S/N) possible to recover diffuse molecular emissions. This yields a synthesized beam of $\sim$2$\arcsec$.3 to 3$\arcsec$ in both lines, which corresponds to $\sim$200 pc, at the Virgo distance of 16 Mpc \citep{yasuda97}. The intensity map was generated using the cube of 10 km~s$^{-1}$ resolution with the smoothing mask in the Astronomical Image Processing System (AIPS), and corrected for primary beam attenuation.

\section{RESULTS} \label{sec:res}

\begin{figure*}[]
\epsscale{1.18}
\plotone{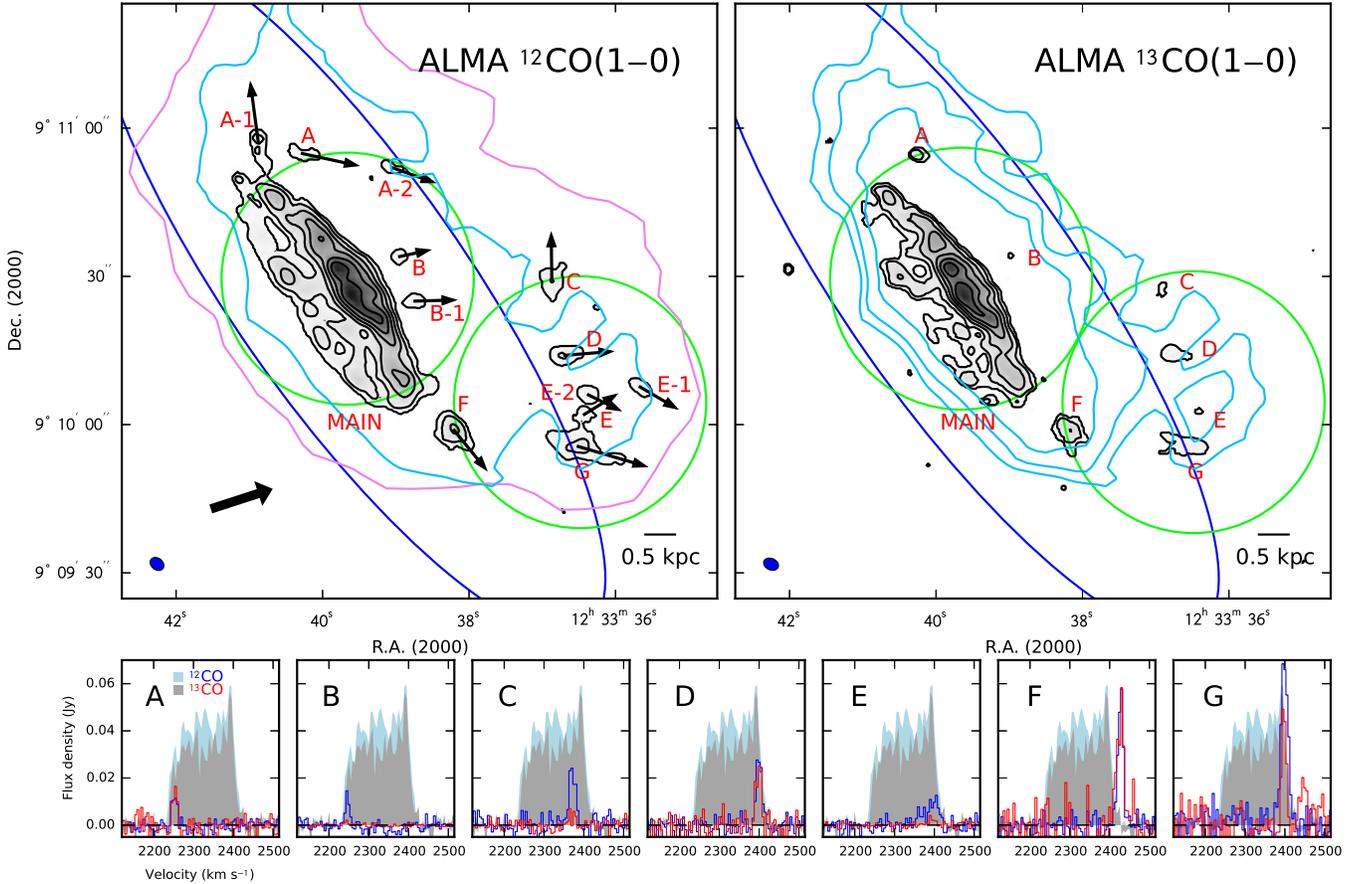}
\caption{\textbf{Top:} ALMA CO intensity maps. The contour levels are (1, 4, 9, 16, 25, 36, 49) $\times$ 0.065 Jy~km~s$^{-1}$~beam$^{-1}$ for $^{12}$CO and (1, 4, 9, 15, 24, 35, 47) $\times$ 0.009 Jy~km~s$^{-1}$~beam$^{-1}$ for $^{13}$CO. The large blue ellipse and green circles represent the $K$-band size and primary beam of ALMA, respectively. The synthesized beam is shown in the bottom-left corner. Small black arrows on the left panel indicate the orientation of each clump, determined by fitting a 2D Gaussian. A thick black arrow shows the ICM wind direction \citep{abramson16}. Magenta and cyan contours represent the {\HI} (0.03 Jy~km~s$^{-1}$~beam$^{-1}$) and single-dish CO (1.7, 3.5, 6.5 Jy~km~s$^{-1}$~beam$^{-1}$) data, respectively. \textbf{Bottom:} CO spectra derived from a tight box around the clumps with a 5 km~s$^{-1}$ resolution. Blue and red lines show the CO flux density of individual clumps, and the light blue and gray backgrounds represent the main disk profile of $^{12}$CO and $^{13}$CO, respectively. The axis corresponds to the scale of $^{12}$CO, whereas $^{13}$CO spectra are magnified by a factor of 8. In the case of the main disk (background), the flux is scaled down by a factor of 0.15 for both lines.  \label{fig:fig1}}
\end{figure*}
\begin{deluxetable*}{cccccccccc}[]
\tablecaption{$^{12}$CO and $^{13}$CO(1--0) properties in NGC~4522 by region\label{tab:table2}}
\tablewidth{0pt}
\tablehead{
\multicolumn{1}{c}{} & \multicolumn{5}{c}{$^{12}$CO(1--0)} & \multicolumn{3}{c}{$^{13}$CO(1--0)} &  \\
\multicolumn{1}{c}{Region} & \multicolumn{1}{c}{Size} & \multicolumn{1}{c}{$v_{^{12}\rm CO}$} & \multicolumn{1}{c}{$\Delta v_{^{12}{\rm CO}}$} & \multicolumn{1}{c}{$S_{^{12}{\rm CO}}$} & \multicolumn{1}{c}{$M_{\rm H_{2}}$} & \multicolumn{1}{c}{$v_{^{13}\rm CO}$} & \multicolumn{1}{c}{$\Delta v_{^{13}{\rm CO}}$} & \multicolumn{1}{c}{$S_{^{13}{\rm CO}}$} & \multicolumn{1}{c}{$R_{12/13}$} \\
& (kpc) & (km~s$^{-1}$) & (km~s$^{-1}$) & (Jy~km~s$^{-1}$) & ($\times$10$^5$\Msun) & (km~s$^{-1}$) & (km~s$^{-1}$) & (Jy~km~s$^{-1}$) &\\
(1) & (2) & (3) & (4) & (5) & (6) & (7) & (8) & (9) & (10)  
}
\startdata
MAIN & 3.99 & 2333.1$\pm$3.0 & 137.8$\pm$7.0 & 64.725$\pm$0.417 & 1289.0$\pm$8.3 & 2333.6$\pm$4.0 & 139.4$\pm$9.3 & 6.148$\pm$0.116 & 10.80$\pm$3.52 \\
A & 0.51 & 2252.3$\pm$0.9 & 15.3$\pm$2.1 & 0.251$\pm$0.026 & 5.0$\pm$0.5 & 2253.6$\pm$1.1 & 13.0$\pm$2.7 & 0.036$\pm$0.004 & 5.42$\pm$1.80 \\
B & 0.25 & 2246.0$\pm$0.7 & 10.5$\pm$1.6 & 0.122$\pm$0.021 & 2.4$\pm$0.4 & 2243.8$\pm$1.2 & 9.6$\pm$2.8 & 0.002$\pm$0.001 & \nodata \\
C & 0.53 & 2371.0$\pm$0.9 & 19.4$\pm$2.1 & 0.523$\pm$0.035 & 10.4$\pm$0.7 & 2369.6$\pm$3.1 & 20.8$\pm$7.2 & 0.007$\pm$0.003 & 10.32$\pm$5.05 \\ 
D & 0.53 & 2399.7$\pm$0.6 & 16.7$\pm$1.5 & 0.448$\pm$0.040 & 8.9$\pm$0.8 & 2398.9$\pm$1.2 & 16.0$\pm$2.7 & 0.036$\pm$0.006 & 8.00$\pm$2.43 \\
E & 0.32 & 2394.2$\pm$2.6 & 40.9$\pm$6.2 & 0.184$\pm$0.040 & 3.7$\pm$0.8 & 2393.6$\pm$2.3 & 17.4$\pm$5.4 & 0.003$\pm$0.002 & \nodata \\
F & 0.75 & 2427.8$\pm$0.4 & 15.3$\pm$0.9 & 1.532$\pm$0.039 & 30.5$\pm$0.8 & 2426.7$\pm$1.2 & 21.6$\pm$2.9 & 0.159$\pm$0.011 & 7.76$\pm$3.31 \\
G & 1.30 & 2399.1$\pm$0.5 & 21.0$\pm$1.2 & 1.364$\pm$0.061 & 27.2$\pm$1.2 & 2397.7$\pm$0.8 & 13.0$\pm$1.8 & 0.068$\pm$0.009 & 9.98$\pm$3.30 \\
\hline
A-1 & 0.60 & 2235.7$\pm$0.9 & 15.4$\pm$2.0 & 0.512$\pm$0.031 & 10.2$\pm$0.6 & \nodata & \nodata & \nodata & \nodata \\
A-2 & 0.34 & 2304.7$\pm$1.1 & 13.3$\pm$2.6 & 0.102$\pm$0.020 & 2.0$\pm$0.4 & \nodata & \nodata & \nodata & \nodata \\
B-1 & 0.36 & 2348.6$\pm$0.9 & 14.1$\pm$2.1 & 0.176$\pm$0.019 & 1.1$\pm$0.4 & \nodata & \nodata & \nodata & \nodata \\
E-1 & 0.39 & 2414.9$\pm$0.7 & 12.8$\pm$1.6 & 0.190$\pm$0.031 & 3.8$\pm$0.6 & \nodata & \nodata & \nodata & \nodata \\
E-2 & 0.36 & 2347.1$\pm$0.9 & 20.6$\pm$2.1 & 0.244$\pm$0.029 & 4.9$\pm$0.6 & \nodata & \nodata & \nodata & \nodata \\
\enddata
\tablecomments{(1) Region, (2) size measured in $^{12}$CO, (3) radial velocity of $^{12}$CO at the Gaussian peak, (4) $^{12}$CO FWHM, (5) $^{12}$CO integrated flux, (6) H$_{2}$ mass estimated from $L_{\rm CO}^{'}$ assuming $\alpha_{\rm CO}$~=~3.2 \Msun~pc$^{-2}$~(K~km~s$^{-1}$)$^{-1}$ \citep{strong96}, (7) radial velocity of $^{13}$CO at the Gaussian peak, (8) $^{13}$CO FWHM, (9) $^{13}$CO integrated flux, (10) mean $^{12}$CO/$^{13}$CO ratio cross each structure measured from the line ratio map (Figure~\ref{fig:fig2}).}
\end{deluxetable*}

Both $^{12}$CO and $^{13}$CO lines were successfully detected, not only inside of the stellar disk but also in the extraplanar space (Figure~\ref{fig:fig1}). The CO line luminosity ($L_{\rm CO}^{'}$) can be measured by the following equation \citep{solomon97}:

\begin{equation}
L^{'}_{\rm CO}=3.25\times10^7 S_{\rm CO}~\nu_{\rm obs}^{-2}~D_{\rm L}^2~(1+\textit{z})^{-3}, 
\label{eqn:lum}
\end{equation}

\noindent in K~km~s$^{-1}$~pc$^{2}$, where $S_{\rm CO}$ is $^{12}$CO(1--0) flux in Jy~km~s$^{-1}$, $D_{\rm L}$ is the luminosity distance in Mpc, $\nu_{\rm obs}$ is the observing frequency in GHz, and \textit{z} is the redshift. To estimate the H$_{2}$ mass ($M_{\rm H_{2}}$), we adopt $\alpha_{\rm CO}$~=~3.2 \Msun~pc$^{-2}$~(K~km~s$^{-1}$)$^{-1}$ \citep{strong96}. For clumps, we measured $S_{\rm CO}$, $L_{\rm CO}^{'}$, and $M_{\rm H_{2}}$ individually. The line ratio of $^{12}$CO to $^{13}$CO ($R_{12/13}$, Figure~\ref{fig:fig2}) was measured for individual clumps as well as the main disk. The result is summarized in Table~\ref{tab:table2}. Note that our estimation for H$_{2}$ mass ($M_{\rm H_{2}}$) does not include helium abundance.

\subsection{Main Disk} \label{subsec:main}
The main molecular gas disk (MAIN in Figure~\ref{fig:fig1}) is found with a similar extent and morphology in two lines. The linewidth and the central velocity of $^{12}$CO and $^{13}$CO are also consistent to within measurement error (Table~\ref{tab:table2}). However, the diffuse emission toward the outer CO disk in the southeast is detected only in $^{12}$CO.  

\begin{figure}
\center
\includegraphics[scale=0.66]{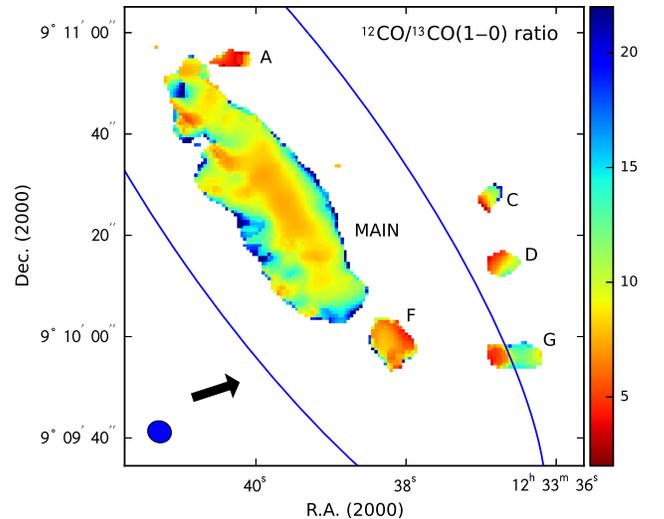}
\caption{Line ratio, $R_{12/13}$ distribution. Both $^{12}$CO and $^{13}$CO maps were imaged at $\sim$4$\arcsec$.5 resolution to increase S/N. The ratio ranges from 3 to 32, with a mean of 10.39 $\pm$ 3.64. The blue ellipse and black arrow are the same as in Figure~\ref{fig:fig1}. \label{fig:fig2}}
\end{figure}

Within the stellar disk, we measure $\sim$$1.3 \times 10^{8}$ \Msun~of $M_{\rm H_{2}}$, including the contribution from clumps (A, B, and F). This corresponds to 87\% of the single-dish flux within the comparable area, assuming a $^{12}$CO (2--1)/(1--0) ratio of $\sim$0.8, as in nearby galaxies \citep{leroy09}. Some diffuse extended CO emission is likely missing from our ALMA data, especially the features extended beyond the maximum recoverable scale ($\sim$21$\arcsec$.5$\approx$1.7 kpc). $R_{12/13}$ across the main disk is 10.80 $\pm$ 3.52 (Figure~\ref{fig:fig2}), which is consistent with the range observed in nearby spiral galaxies \citep[e.g.,][]{paglione01, vila15}. The main CO disk in the ALMA data appears to be more extended in the east, i.e., opposite the ICM wind direction. However, this is only due to the projection, and the outermost CO disk is more compressed on this side, as the single-dish data clearly show (top right of Figure~\ref{fig:fig1}).

\subsection{Clumps} \label{subsec:clump}
Within two pointings, about 8\% of the total CO flux has been detected in clumps, about half of which are found in the extraplanar space. As shown in Figure~\ref{fig:fig1}, clumps are nominated in descending order of declination, from A to G. Those detected only in $^{12}$CO are numbered under the same letter of the nearest clump seen in both lines, and in the order of decreasing declination, starting from 1. Because of the superior sensitivity of the ALMA, most of the extraplanar clumps (C, D, E, and G) have been also detected in $^{13}$CO, which is the first extraplanar $^{13}$CO detection ever reported in a galaxy undergoing ram pressure stripping.

\begin{figure}
\center
\includegraphics[scale=0.645]{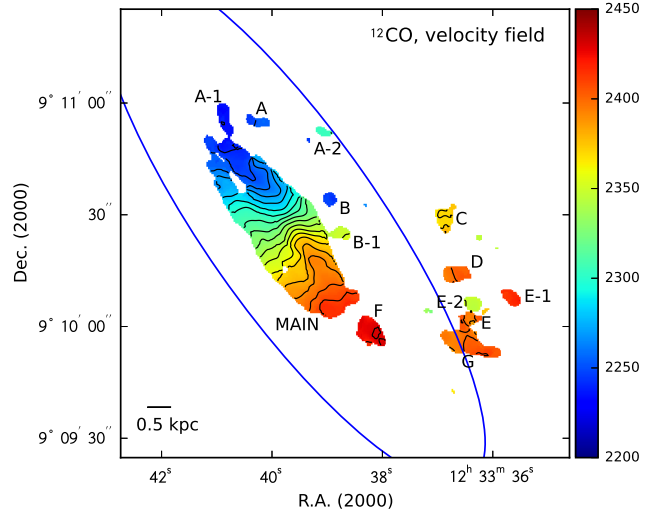}
\caption{Velocity field map of $^{12}$CO. Contours are drawn from 2240~km~s$^{-1}$ to 2430~km~s$^{-1}$ with a 10~km~s$^{-1}$ interval. \label{fig:fig3}}
\end{figure}

The H$_{2}$ mass of clumps ranges from $10^{5}$~\Msun~to $3 \times 10^{6}$~\Msun, which is comparable to Galactic giant molecular clouds (GMCs). The central velocity and linewidth of individual clumps (Table~\ref{tab:table2}) have been measured by fitting a single 1D Gaussian function to each spectrum shown in the bottom panels of Figure~\ref{fig:fig1}. Clumps are more compact in $^{13}$CO than in $^{12}$CO, but the velocity and linewidth generally show good agreement between the two lines (Table~\ref{tab:table2}). The velocity of the extraplanar clumps (C, D, G, and/or E) does not exceed the highest velocity end on the same side (southwest) of the main disk ($\sim$2420 km~s$^{-1}$), with a mean of 2388 km~s$^{-1}$ (Figure~\ref{fig:fig3}). That is, for their galactocentric radii, these clumps are shifted to the lower-velocity side, toward the cluster mean ($\sim$1100 km~s$^{-1}$, \citealt{mei07}). The velocity offset of Clump B from the main disk is also interesting, as seen in Figure~\ref{fig:fig1} and \ref{fig:fig3}. Although it is located near the galaxy's center, the velocity is offset from the center of the main gas disk by $\sim$90 km~s$^{-1}$, to the lower velocity, i.e., toward the cluster mean. We discuss its implication in more detail in the following section.

We measured $R_{12/13}$ for the clumps detected in 5$\sigma$ or above in both lines. $R_{12/13}$ is generally lower compared to the main disk, but not significantly. More intriguingly, the clumps outside the disk mostly show an increasing gradient from east to west, which will also be discussed further in the following section.

\section{DISCUSSION} \label{sec:disc}
\subsection{Characteristics of Extraplanar Clumps} \label{subsec:rps}

The extraplanar CO gas discovered in the previous single-dish observation has been resolved into a number of clumps. Within the second ALMA pointing, $\sim$20\% of the single-dish flux is missing, yet the clumps occupy only $\sim$26\% of the extraplanar gas in area. This implies that the extraplanar molecular gas in this galaxy has a clumpy nature, and the clumps are less likely to be the local peaks of more extended large-scale emission. Instead, these clumpy cold molecular features might be embedded and shielded by an envelope of warmer H$_{2}$ gas, which was confirmed in infrared \citep{wong14}.

By fitting a single 2D Gaussian, we measured the orientation of individual clumps. Most clumps are not randomly oriented but elongated along the east-west direction. For 75\% of the clumps (83\% among the extraplanar clumps), the position angle ranges from 234 degrees to 302 degrees, roughly pointing the direction of the projected ICM wind ($\sim$288 degrees, \citealt{abramson16}). The elongation of ram-pressure-stripped features in the same direction as the ICM wind is commonly observed in a range of wavelengths, and also in simulations \citep{tonnesen10, abramson14}. 

Compared to the main disk at the same radii, the central velocities of clumps are generally offset to the lower side, toward the Virgo center ($\sim$1100 km~s$^{-1}$, \citealt{mei07}), as expected for the gas stripped due to ram pressure \citep[e.g.,][]{kenney04}. In addition to the overall molecular gas morphology seen in the previous single-dish observations, all of these characteristics of clumps strongly support the impact of ICM winds on the molecular gas.

The distribution of the line ratio between $^{12}$CO and $^{13}$CO, $R_{12/13}$, across the extraplanar clumps is also intriguing. $R_{12/13}$ within the main disk falls well within the range observed in nearby galaxies \citep{paglione01, vila15}. Meanwhile, $R_{12/13}$ of our extraplanar clumps is comparable to the disk ratio. For comparison, the intergalactic molecular gas pulled out during galaxy--galaxy interactions shows much higher $R_{12/13}$ (e.g., $\sim$50 in the bridge of UGC 12914/5; \citealt{braine03}, $>$25 in the tidal tail of Stephan's Quintet; \citealt{lisenfeld04}). The high $R_{12/13}$ in those cases may indicate that the gas density has declined as it becomes optically thin during dispersal into the surrounding space \citep[e.g.,][]{konig16}. On the other hand, our extraplanar molecular clumps are located in a high-pressure environment, which potentially helps the gas stay dense by compression \citep{alatalo15}. Alternatively, chemical fractionation of CO, and hence the enhancement of $^{13}$CO abundance, can also result in relatively low $R_{12/13}$ \citep{watson76,szucs14}.

Intriguingly, $R_{12/13}$ of individual extraplanar clumps (C, D, and G in Figure~\ref{fig:fig2}) shows an increasing gradient from east to west. The gradient of the line ratio suggests a variation of optical depth, indicating the change in physical conditions within each extraplanar molecular clump \citep[e.g.,][]{aalto10}. Assuming that the physical and chemical conditions, such as temperature, dispersion, and abundance of the gas, do not vary significantly within each clump, this gradient may indicate a higher density in the east than in the west \citep[e.g.,][]{sakamoto97, konig16}. Because of the high resolution of the ALMA, we are able to see the gas being compressed in the ICM wind front \citep{nehlig16, lee17}, even at this small scale.

~~~~   
  
~~~~ 

~~~~      
\subsection{Origin of Extraplanar Clumps} \label{subsec:ori}

Among the galaxies showing clear signs of active ram pressure stripping, only a few cases have been found to date with molecular gas outside of the main disk \citep[e.g.,][]{jachym14, jachym17, verdugo15}. However, there is no consensus on its origin, or whether those features are the stripped molecules from the disk or formed in situ from stripped diffuse atomic gas \citep{jachym14, jachym17, moretti18}. 

To examine the origin of NGC~4522's extraplanar CO clumps, we have compared various timescales involved with molecular gas stripping and formation processes. Naively thinking, if it takes too long to form molecules outside of the galactic disk, extraplanar CO clumps could be simply the result of molecular gas stripping.

At {\HI} truncation radii and beyond, it should be feasible to strip a lower-density component of the extended molecular features or the outer part of molecular clouds. We do not include the possibility of stripping molecular gas as clouds, as their cores are usually too dense to be stripped by realistic ram pressure \citep[e.g.,][]{abramson14}. By measuring the youngest stellar age at the {\HI} and H$\alpha$ truncation radius of NGC~4522,  \cite{crowl08} suggested that it is $\sim$100 Myr ago when {\HI} gas was stripped and thus star formation was shut down, which is also consistent with the timescale constrained using the simulated model by \cite{vollmer06}. Therefore, $\sim$100 Myr can be taken as an upper limit of the molecular gas stripping timescale in the case of NGC~4522.

Meanwhile, if molecules have been formed from stripped atomic gas, the formation timescale should be added to the stripping timescale mentioned above. The timescale for the transition from atomic to molecular hydrogen is $\simeq$10 Myr in typical interstellar clouds \citep[e.g.,][]{goldsmith07}. In addition, considering the clumpy nature of the extraplanar molecular gas in NGC~4522, one should consider how long it would take for molecules to form clouds. $M_{\rm H_{2}}$ of our clumps is comparable to those of Galactic GMCs, for which is estimated to require a few tens of Myr to form \citep[e.g.,][]{inutsuka15}.

However, the linewidth and size of our clumps are larger than those of Galactic GMCs, which implies that these extraplanar molecular clumps are unlikely to be virialized. Therefore, these clumps should be younger than a few tens of Myr, which is comparable to the uncertainty of the gas stripping timescale.

Conclusively, the timescales for different scenarios, (1) diffuse molecular gas stripping (recrumpled as clouds afterwards) versus (2) diffuse atomic gas stripping followed by molecule and cloud formation, are not significantly distinct, and we cannot uniquely constrain the origin of these extraplanar molecular clumps based on the timescales alone. Alternatively, both processes might be responsible, as suggested by \cite{jachym14}. Indeed, based on a simple \cite{gunn72}'s prescription, molecular gas stripping is not impossible in the case of NGC~4522 \citep{abramson14, lee17}. At the same time, the column density of the extraplanar {\HI} gas around our CO clumps  exceeds the critical density required to form molecular hydrogen \citep[e.g.,][references therein]{burkhart16}; therefore, both scenarios are quite feasible.

\subsection{Role of Ram Pressure Stripping in the Chemical Enrichment of Halo Gas and ICM} \label{subsec:chem}

No matter how molecules ended up in the extraplanar space, the presence of $^{13}$CO outside of the stellar disk strongly suggests that at least $^{13}$C originated from the disk. Unlike $^{12}$C, which can be synthesized and donated to the interstellar field by massive stars within a relatively short timescale ($\sim$a few tens Myr), the most dominant donor of $^{13}$C is the asymptotic giant branch (AGB) stage of intermediate-mass stars \citep{milam05,jimenez17}. Recalling the fact that it takes $\sim$100 Myr or longer for intermediate-mass stars to evolve into AGBs, it should take much longer for molecular gas to form stars that can donate $^{13}$C into the surrounding medium. 

In our case, the extraplanar cool gas has H$\alpha$ counterpart, but not far ultraviolet, indicating that stars have been newly formed within the last tens of Myr \citep{kennicutt98} outside the disk. Thus, $^{13}$C is unlikely to come from the newly formed stars in the halo, but must have come from the disk, either as $^{13}$CO molecules or synthesized to $^{13}$CO through a chemical reaction over a few Myr \citep{glover10,szucs14} after being stripped in the atomic phase. This is also supported by the similar oxygen abundances in the extraplanar star forming region and in the disk of NGC~4522 \citep{stein17}.

Several mechanisms, such as galactic winds, active galactic nuclei (AGN) outflows, ram pressure stripping, and galaxy--galaxy interactions, have all been suggested as the driver of the chemical evolution of the ICM \citep[e.g.,][]{kapferer05,domainko06,simionescu08}. Our $^{13}$CO data provide observational evidence that a ram pressure stripping event can transport heavy elements from the disk to the extraplanar space by momentum transfer. 

If the ram pressure is strong enough to completely strip the heavy elements from the galaxy, it will be able to chemically enrich the ICM \citep{domainko06,urban11}. However, whether this will be the case strongly depends on the orbital motion of the galaxy and/or the time evolution of the surrounding ICM density. In fact, other heavy elements, such as O and N, have previously been found quite far from a galaxy undergoing ram pressure stripping (e.g., along a few tens kpc-long tail of ESO 137-001; \citealt{fossati16}), but compared to $^{13}$C, their formation timescale is very short and it is not guaranteed that they originate from the disk. In order to probe how far stripped metals can reach and whether they contribute to the enrichment of the ICM, more chemical studies will need to be carried out along with the ram-pressure-stripped gas tails.

\acknowledgments
This Letter makes use of the following ALMA data: ADS/JAO.ALMA\#2015.1.01341.S. ALMA is a partnership of ESO (representing its member states), NSF (USA) and NINS (Japan), together with NRC (Canada), MOST and ASIAA (Taiwan), and KASI (Republic of Korea), in cooperation with the Republic of Chile. The Joint ALMA Observatory is operated by ESO, AUI/NRAO and NAOJ. Support for this work was provided by the National Research Foundation of Korea to the Center for Galaxy Evolution Research (No. 2010-0027910) and NRF grant No. 2015R1D1A1A01060516 and 2018R1D1A1B07048314. We acknowledge the usage of the HyperLeda database (\url{http://leda.univ-lyon1.fr/}). We acknowledge Bon-Chul Koo, O. Ivy Wong, and Susanne Aalto for useful comments and discussion, and the Korea ARC team for their help with the data.

\end{document}